\documentclass{article}

\usepackage{amsmath,amssymb,amsthm}
\usepackage{tikz}
\usepackage{color}
\usepackage[toc]{appendix}
\usepackage{graphicx}

\usepackage{parskip}
\usepackage{float}
\usepackage{chngpage}
\usepackage{calc}
\usepackage{bigints}
\usepackage{array}
\usepackage{booktabs}
\usepackage{rotating}
\usepackage{multirow}
\usepackage{adjustbox}
\usepackage{tabularx}
\usepackage{verbatim}
\usepackage{mathtools}
\usepackage{ragged2e}
\usepackage[makeroom]{cancel}
\usepackage{caption}
\usepackage{hyperref}
\usepackage{caption}
\usepackage{subcaption}
\usepackage{appendix}
\usepackage{pgfplots}
\pgfplotsset{compat=1.16}
\usepackage[english]{babel}
\usepackage{hyphenat}
\usepackage[makeindex]{imakeidx}

\usepackage[sort&compress,round,comma,authoryear]{natbib}

\usepackage{amsmath}
\DeclareMathOperator*{\argmax}{arg\,max}

\usetikzlibrary{datavisualization}
\usetikzlibrary{matrix}
\usetikzlibrary{datavisualization.formats.functions}

\newtheorem{theorem}{Theorem}[section]

\newtheorem{remark}[theorem]{Remark}
\newtheorem{assumption}[theorem]{Assumption}

\makeatletter
\def\section{\@startsection {section}{1}{\z@}{3.25ex plus 1ex minus
		.2ex}{1.5ex plus .2ex}{\large\bf}}
\def\subsection{\@startsection{subsection}{2}{\z@}{3.25ex plus 1ex minus
		.2ex}{1.5ex plus .2ex}{\normalsize\bf}}
\@addtoreset{equation}{section} 
\makeatother

\title{A Novel Theoretical Framework for Exponential Smoothing}

\author{ Enrico Bernardi \thanks{Dipartimento di Scienze Statistiche Paolo Fortunati, Università di Bologna, Bologna, Italy. \textbf{e-mail}: enrico.bernardi@unibo.it} \and
Alberto Lanconelli\thanks{Dipartimento di Scienze Statistiche Paolo Fortunati, Università di Bologna, Bologna, Italy. \textbf{e-mail}: alberto.lanconelli2@unibo.it} \and Christopher S. A. Lauria\thanks{Dipartimento di Scienze Statistiche Paolo Fortunati, Università di Bologna, Bologna, Italy. \textbf{e-mail}: christopher.lauria2@unibo.it}}

\date{\today}

\begin{document}
	
\maketitle

\begin{abstract}
Simple Exponential Smoothing is a classical technique used for smoothing time series data by assigning exponentially decreasing weights to past observations through a recursive equation; it is sometimes presented as a rule of thumb procedure.
We introduce a novel theoretical perspective where the recursive equation that defines simple exponential smoothing occurs naturally as a stochastic gradient ascent scheme to optimize a sequence of Gaussian log-likelihood functions.
Under this lens of analysis, our main theorem shows that -in a general setting- simple exponential smoothing converges to a neighborhood of the trend of a trend-stationary stochastic process. This offers a novel theoretical assurance that the exponential smoothing procedure yields reliable estimators of the underlying trend shedding light on long-standing observations in the literature regarding the robustness of simple exponential smoothing.
\end{abstract}
	
	
	
\allowdisplaybreaks
	
\bigskip
	
\section{Introduction}\label{s1}
Formally defined in \citet{brown1956exponential}, and then further developed in \citet{brown1959statistical} and \citet{brownfore_pre}, Simple Exponential Smoothing (SES) was initially introduced to generate forecasts for time series data by giving more weight to recent observations while exponentially diminishing the influence of older data. Mathematically, the SES procedure is succinctly captured by a straightforward recursive expression
\begin{align}   \label{ses}
S_{t+1} = S_t + \alpha (X_{t+1} - S_t),
\end{align}
where $X_{t+1}$ denotes the observation at time $t+1$, $ 0<\alpha< 1$ is the so called \emph{smoothing parameter} and $S_0 \in \mathbb{R}$ is a chosen initial value.  \\
SES enjoyed success in multiple applications, \citet{winters1960forecasting, makridakis1982accuracy,Gross}, and was generalized in a number of ways; a taxonomy can be found in \citet{hyndman2002state, taylor2003exponential} and \citet{gardner2006exponential}. A remarkable property of SES is its robustness with regards to the assumed data generating process, a quality observed in several publications, \citet{cohen1963note,cox1961prediction,tiao1993robustness}. Furthermore, \citet{bossons1966effects} showed that  SES tends to be largely unaffected by specification errors, while \citet{hyndman2001s} explains how ARIMA model selection errors can significantly amplify mean square deviations when compared to SES.\\ 
Despite its success in empirical settings SES has sometimes been criticized for being "an \textit{ad hoc} procedure with no statistical rationale", as recounted in \citet{gardner2006exponential}. The main procedure used to theoretically justify exponential smoothing methods has been to define statistical models for which the optimal forecasts are equivalent to those obtained from the exponential smoother. In \citet{muth1960optimal} two models are provided for which the optimal forecasts are given by the SES equation with appropriately chosen smoothing parameter. More recently, \citet{ord1997estimation} defined a class of state-space models with a single source of error for which the forecasts are given by the Holt-Winters exponential smoother \citet{winters1960forecasting}; subsequently, \citet{hyndman2002state} extended the work of \citet{ord1997estimation} to include numerous other exponential smoothing procedures. For a clear and comprehensive presentation of the state space framework for exponential smoothing see \citet{hyndman2008forecasting}. \\
In this paper, we will establish a novel bound on the estimates generated by SES that holds in a surprisingly general setting. Our approach involves a clear statistical rationale, derived from the classical theory of maximum likelihood estimation, that sheds light on longstanding observations in the literature concerning the robustness properties of SES.
The paper is organized as follows: in subsection \ref{sub1} we show how SES can be viewed as time varying stochastic gradient ascent scheme. In subsection \ref{sub2} we formally state our main result and in section \ref{sec2} we provide some relevant simulations. Section \ref{sec3} will be dedicated to the proof of the main result.

\section{Exponential Smoothing as Stochastic Gradient Ascent} \label{sub1}

A trend-stationary stochastic process is a stochastic process from which a function of time can be subtracted, leaving a stationary process. Formally, a stochastic process $\{X_t \}_{t \in \mathbb{N}}$, is said to be trend stationary if
\begin{align} \label{trend-sta}
	X_t = m_t^{\star} + \epsilon_t,\quad t\in\mathbb{N},
\end{align}
for some sequence of real numbers $\{m_t^{\star}\}_{t\in\mathbb{N}}$, called \emph{trend}, and some weakly stationary sequence $\{ \epsilon_t \}_{t \in \mathbb{N}}$ with $\mathbb{E}[\epsilon_t] = 0 $ and $\mathtt{cov}(\epsilon_t, \epsilon_s ) \coloneqq \gamma(t, s) = \gamma(t-s)$ for all $t,s \in \mathbb{N}$; here, $\gamma: \mathbb{Z} \rightarrow \mathbb{R}$ is an even positive definite function. 
In this setting, SES has been used as a method for estimating the trend $\{m_t^{\star}\}_{t\in\mathbb{N}}$ through the observed values $\{X_t\}_{t\in\mathbb{N}}$, \citet{brockwell2009time}. \\
We will now show how SES \eqref{ses} can be cast as a stochastic gradient ascent algorithm employed for solving an optimization problem.
First we relabel the SES recursive equation \eqref{ses} by defining $\hat{m}_{t+1} \coloneqq S_t$, for all $t\geq 1$, resulting in
\begin{align}
\label{SES intro*}
	\hat{m}_{t+1}=(1-\alpha)\hat{m}_{t}+\alpha X_{t},\quad t\in\mathbb{N}
\end{align}
Now the sequence $\{\hat{m}_t\}_{t\in\mathbb{N}}$ is simply a delayed version of \eqref{ses}: $\hat{m}_{t+1}$ is an estimate for $m^*_t$ and not for $m^*_{t+1}$. Taking as initial value the common choice $ \hat{m}_1 = X_1$ we then rewrite equation \eqref{SES intro*} as 
\begin{align} 
\label{SES intro***}
		\hat{m}_{t+1}=\hat{m}_{t}+\alpha (X_{t}-\hat{m}_t) = \hat{m}_{t}+\tilde{\alpha} \frac{d}{d\hat{m}}\left[\ln\left(\frac{1}{\sqrt{2\pi\gamma(0)}}e^{-\frac{(X_{t}-\hat{m})^2}{2\gamma(0)}}\right)\right]\Big|_{\hat{m}=\hat{m}_t},\quad t\in\mathbb{N}  
\end{align}
with $\tilde{\alpha}:=\alpha\gamma(0)$. If the sequence $\{\epsilon_t\}_{t\in\mathbb{N}}$ in \eqref{trend-sta}, and hence  $\{X_t\}_{t\in\mathbb{N}}$, is Gaussian (such assumption can be relaxed, see Remark \ref{non gaussian} below), then the expression between square brackets in \eqref{SES intro***} coincides with the log-likelihood of $X_t$ evaluated at $\hat{m}_t$; from this point of view the scheme \eqref{SES intro***} takes the form of a stochastic gradient ascent algorithm with constant learning rate $\tilde{\alpha}$.
Notice however that, at each time $t$, the stochastic gradient employed belongs to a different Gaussian: this type of algorithm is employed in the optimization literature to track an optimum that moves through time. \\
Tracking a sequence of optima through time is a problem that has already been considered in a variety of settings, see for instance \citet{Polyak, Popkov2005, Zinkevich, Bubeck,Cao}. Using the nomenclature of \citet{simonetto2020time}, these types of problems have been called  \emph{time-varying} optimization problems since their aim is to find the optima $t \mapsto   m^*_t \coloneqq \argmax f(t,m) $ of a sequence of objective functions $\{ f(t,\cdot) \}_{t \in \mathbb{N}}$. A computationally efficient manner to track the optima is via recursive schemes that utilize the gradient of the sequence $\{ f(t,\cdot) \}_{t \in \mathbb{N}}$ \citet{Simonetto, Simonetto1}. The simplest prototypical algorithmic procedure that has been considered is to employ a standard gradient ascent scheme where at the $t$-th iteration the gradient is computed with respect to the $t$-th objective function, in formulas:
\begin{align}   \label{gdf}
	\hat{m}_{t+1} = \hat{m}_t + \alpha \frac{d}{d\hat{m}} f(t,\hat{m})\Big|_{\hat{m}=\hat{m}_t},\quad t\in\mathbb{N},  
\end{align} 
where $\alpha> 0$ is the so-called \emph{learning rate}. If the distance between optimizers at subsequent times is uniformly upper bounded, then running the recursive equation \eqref{gdf} will track the solution trajectory up to a neighborhood \citet{simonetto2020time}. We can now observe that taking
\begin{align} \label{tvop}
f(t, \hat{m}) \coloneqq \mathbb{E}\left[\ln\left(\frac{1}{\sqrt{2\pi\gamma(0)}}e^{-\frac{(X_{t}-\hat{m})^2}{2\gamma(0)}}\right)\right]
\end{align}
equation \eqref{gdf} becomes 
\begin{align*}  
	\hat{m}_{t+1} = \hat{m}_t + \alpha \frac{d}{d\hat{m}} \mathbb{E}\left[\ln\left(\frac{1}{\sqrt{2\pi\gamma(0)}}e^{-\frac{(X_{t}-\hat{m})^2}{2\gamma(0)}}\right)\right]  \Big|_{\hat{m}=\hat{m}_t},\quad t\in\mathbb{N},  
\end{align*} 
that through a stochastic approximation of the expected value \citet{Robbins} is analogous to equation \eqref{SES intro***}. The trend $\{m_t^*\}_{t\in\mathbb{N}}$  is the solution to the optimization problem defined by the sequence of objective functions  $$ \left\{ \mathbb{E}\left[\ln\left(\frac{1}{\sqrt{2\pi\gamma(0)}}e^{-\frac{(X_{t}-\hat{m})^2}{2\gamma(0)}}\right)\right] \right\}_{t\in\mathbb{N}} .$$
Therefore, the SES method \eqref{SES intro*} for trend estimation is equivalent to a time varying stochastic gradient ascent algorithm that tracks the maxima of the log-likelihoods associated with the observations \eqref{trend-sta}.

\begin{remark}  If the trend $\{m_t^*\}_{t\in\mathbb{N}}$ is constant, i.e. $m_t^*=m^*$ for all $t\in\mathbb{N}$, we have a single objective function to maximize, as is canonical in the classical theory of maximum likelihood estimation. In this case one can simply use the arithmetic mean to estimate $m^*$.
\end{remark}

\begin{remark}\label{non gaussian}
	In the case where the sequence $\{\epsilon_t\}_{t\in\mathbb{N}}$ is not Gaussian, and thus for $t\in\mathbb{N}$ the random variable $X_t$ has probability density function $x\mapsto p(\cdot-m_t)$ for some non negative $p$ with $\int_{\mathbb{R}}p(x)dx=1$ and absolute maximum at the origin,  we can still run a stochastic gradient ascent algorithm to track the trend $\{m_t\}_{t\in\mathbb{N}}$ as 
	\begin{align*} 
	\hat{m}_{t+1} & = \hat{m}_t + \alpha \frac{d}{d\hat{m}} \ln p(X_t-\hat{m}) \Big|_{\hat{m}=\hat{m}_t}. 
	\end{align*}
	Now, by an application of the Mean Value Theorem we can write
	\begin{align}\label{c} 
	\hat{m}_{t+1} & = \hat{m}_t + \alpha \frac{d}{d\hat{m}} \ln p(X_t-\hat{m}) \Big|_{\hat{m}=\hat{m}_t}\nonumber\\
	& = \hat{m}_t + \alpha \frac{d}{d\hat{m}} \ln p(X_t-\hat{m}) \Big|_{\hat{m}=X_t}+\alpha \frac{d^2}{d\hat{m}^2} \ln p(X_t-\hat{m}) \Big|_{\hat{m}=\xi_t}(\hat{m}_t-X_t)\nonumber\\
	& = \hat{m}_t + \alpha \frac{d^2}{d\hat{m}^2} \ln p(X_t-\hat{m}) \Big|_{\hat{m}=\xi_t}(\hat{m}_t-X_t);
	\end{align}
	here we have utilized the fact that $X_t$ is the maximum likelihood estimator of $m_t^*$ while $\xi_t$ belongs to the segment $[ X_t,  \hat{m}_t ] $. If classical conditions to prove convergence of gradient ascent apply (see for instance \citet{Polyak}), i.e. the function $ \hat{m} \rightarrow -\ln p( x - \hat{m} )$ is strongly convex with convexity constant $\ell$ and has Lipschitz continuous derivative with constant $L$ (both constants $\ell,L$  being independent of $x$), then we can bound the second derivative as  $ \ell \le -\frac{d^2}{d\hat{m}^2} \ln p(X_t-\hat{m}) \Big|_{\hat{m}=\xi_t} \le L $. 
	Therefore, replacing $-\frac{d^2}{d\hat{m}^2} \ln p(X_t-\hat{m}) \Big|_{\hat{m}=\xi_t}$ with an appropriately chosen constant $ \rho\in[\ell,L]$ one reduces equation \eqref{c} to
	\begin{align*} 
	\hat{m}_{t+1} = \hat{m}_t + \alpha \rho ( X_t- \hat{m}_t ),
	\end{align*}
	that, up to a reparametrization of the learning rate, coincides with \eqref{SES intro*}. This remark suggests that viewing the SES procedure as a stochastic gradient ascent algorithm can be appropriate even outside the Gaussian case.
\end{remark}

\begin{remark} \label{qcf}
Equation \eqref{SES intro*} may also be viewed as a time-varying stochastic gradient descent algorithm used to minimize the quadratic cost functions 
	\begin{align*}
	\hat{m} \mapsto \frac{1}{2}\mathbb{E}[ (  \hat{m} - X_t   )^2 ],\quad t\in\mathbb{N}
	\end{align*}
	where the gradient of the objective function is replaced by its random selection $(\hat{m}_t-X_t)$. These objective functions will for each time $t$ achieve their minimum at the desired $m_t^*=\mathbb{E}[X_t]$. Under these lens SES may be understood as a time varying M-estimator, \citet{godambe1991estimating, hayashi2011econometrics}, obviating the necessity for an a priori assumption regarding the Gaussian distribution of the observations.
\end{remark}

The aim of the present paper is to prove novel quantitative estimates for the performance of SES \eqref{SES intro*} by utilizing techniques from the optimization literature. We will show that, under only two assumptions, SES tracks the trend $\{m_t^*\}_{t\in\mathbb{N}}$ of a trend-stationary stochastic process $\{X_t\}_{t\in\mathbb{N}}$ up to a neighborhood. Our result provides novel insights on how to choose the smoothing parameter, and showcases how SES performs well in a wide variety of settings shedding light on the robustness properties of SES.  
A similar approach, where the observations are independent, but their distribution belongs to a wider class than the one studied here can be found in \citet{lanconelli}, \citet{onlinel} ; there the authors prove that after a certain number of iterations the stochastic gradient method, used to estimate a time-varying parameter, is able to improve on the naive estimator given by maximizing the single observation log-likelihood. 

\section{Assumptions and statement of the main result} \label{sub2}

We now collect the assumptions needed to prove our main theorem.

\begin{assumption} [Trend-stationarity of the observations] \label{Stationarity}
	The stochastic process $\{X_t \}_{t \in \mathbb{N}}$ is trend-stationary with
\begin{align} \label{trendsta}
X_t = m_t^{\star} + \epsilon_t,\quad t\in\mathbb{N};  
\end{align}
the trend $\{m_t^{\star}\}_{t\in\mathbb{N}}$ is a sequence of real numbers while $\{ \epsilon_t \}_{t \in \mathbb{N}}$ is a weakly stationary sequence with $\mathbb{E}[\epsilon_t] = 0 $ and $\mathtt{cov}(\epsilon_t, \epsilon_s ) \coloneqq \gamma(t, s) = \gamma(t-s)$ for all $t,s \in \mathbb{N}$; the function $\gamma: \mathbb{Z} \rightarrow \mathbb{R}$ is even and positive definite.
\end{assumption} 

The next assumption concerns the evolution of the trend $\{m_t^{\star}\}_{t\in\mathbb{N}}$. 

\begin{assumption}[Lipschitz continuity of the trend] \label{ltvp}
There exists a positive constant $K$ such that
\begin{align*}
| m_{t+1}^{\star} - m_t^{\star} | \le K\quad\mbox{ for all $t\in\mathbb{N}$}.
\end{align*}
\end{assumption} 

Assumption \ref{ltvp} is canonical in the time-varying optimization literature, see for example \citet{simonetto2020time, Cao, Wilson} and \citet{onlinel}.: a control on the behavior of the sequence of optima is needed to be able to shadow it. 

We can now state our main theorem.

\begin{theorem} \label{t1} 
Let Assumptions \ref{Stationarity} and \ref{ltvp} hold. Then, running equation 
\begin{align}
	\begin{cases}\label{SES th}
		\hat{m}_{t+1}=(1-\alpha)\hat{m}_{t}+\alpha X_{t},\quad t\in\mathbb{N}; \\
		\hat{m}_1=X_1,
	\end{cases}
\end{align}
with $\alpha\in ]0,1[$ we obtain
\begin{align}\label{limsup}
 \limsup_{t \rightarrow \infty} \mathbb{E}[( \hat{m}_{t+1}-m_t^{\star} )^2 ]  \le  \frac{ \alpha }{ 2-\alpha } \gamma(0) + \frac{2\alpha}{2 - \alpha}  \sum_{k\geq 1} \gamma(k) (1-\alpha)^k +\frac{(1-\alpha)^2}{\alpha^2} K^2.
\end{align}
\end{theorem}
The proof of Theorem \ref{t1}, presented in Section \ref{sec3}, ensues through the application of methodologies conventionally employed in the optimization literature for establishing the convergence of stochastic gradient algorithms.
Theorem \ref{t1} elucidates the conditions under which SES is poised to be efficacious, thereby offering valuable insight regarding the potential utility of employing SES within different applied settings. Moreover, the generality of the conditions under which Theorem \ref{t1} holds explain the robustness properties of SES; even in the case of model misspecification (when SES does not represent the optimal estimator with respect to a certain data generating process) it will still converge under the Lipschitz trend Assumption \ref{ltvp} to a neighborhood of the true trend. \\
We also observe that as $\alpha$ approaches zero the first two terms on the right-hand side of inequality \eqref{limsup} converges to zero, whereas the third term diverges. This phenomenon arises due to the inadequacy of a small $\alpha$'s to effectively capture the evolving dynamics of the sequence $\{m_t^{\star}\}_{t\in\mathbb{N}}$. Consequently, selecting an $\alpha$ that diminishes with time, in accordance with the conventional approach in the optimization literature dealing with static optima, proves to be less efficacious in improving convergence.

\begin{remark}\label{contribution}
The right-hand side of inequality \eqref{limsup} clarifies the contribution of the assumptions to the performance of the method. Given the learning rate $\alpha$ we have: 
\begin{align*}
\underbrace{ \frac{ \alpha }{ 2-\alpha } \gamma(0) }_\text{Variance of the observations}\quad + \quad\quad\underbrace{\frac{2\alpha}{2 - \alpha}  \sum_{k=1}^{\infty} \gamma(k) (1-\alpha)^k }_\text{Correlation structure}     \quad +\quad\underbrace{\frac{(1-\alpha)^2}{\alpha^2} K^2}_\text{Dynamics of the trend}.
\end{align*}
The first term increases with the variance of the observations, the second one is ruled by the covariances between observations at different times while the third is due to the time evolution of the trend. When one possesses prior knowledge regarding the covariance structure of the underlying process and the temporal dynamics of the trend, the information can be effectively leveraged to optimize the selection of a suitable value for the parameter $\alpha$.
\end{remark}

\section{Simulations} \label{sec2}

In this section we present simulations showcasing how SES converges to a neighborhood of the trend $\{m_t^{\star}\}_{t\in\mathbb{N}}$ of a trend-stationary stochastic process $\{X_t\}_{t\in\mathbb{N}}$: we consider the cases where the stationary process $\{\epsilon_t\}_{t\in\mathbb{N}}$ is given by a moving avarage process $\mathtt{MA(1)}$ and an autoregressive model $\mathtt{AR(1)}$.

\subsection{The moving average case}

\begin{figure}[h]
	\centering
	\begin{subfigure}[t]{0.49\textwidth}
		\includegraphics[width=\linewidth]{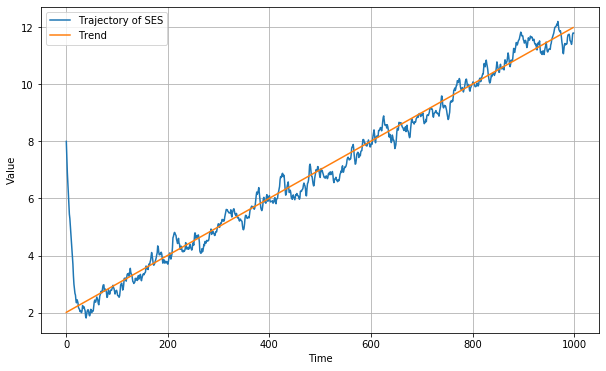}
		\caption{}
		\label{fig:1a}
	\end{subfigure}
	\begin{subfigure}[t]{0.49\textwidth}
		\includegraphics[width=\linewidth]{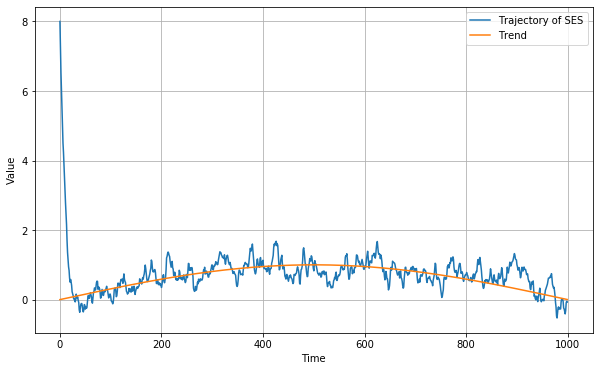}
		\caption{}
		\label{fig:1b}
	\end{subfigure}

\caption{The evolution of $\{m_t^{\star}\}_{t\in\mathbb{N}}$ in subfigure \ref{fig:1a} is given by the equation $m_t^{\star} = m_{t-1}^* + 0.1$, starting from $m_0^* = 2$, while $a = 2$. The evolution of $\{m_t^{\star}\}_{t\in\mathbb{N}}$ in subfigure \ref{fig:1b} is given by the equation $m_t^{\star} = \sin( \pi t / 1000)$, starting from $t = 0$, while $a = 2$. The SES in both subfigures is iterated $1000$ times with $\hat{m}_0= 8$ and $\alpha = 0.1$.}
\label{fig1}
\end{figure}

We simulate observations of the form
\begin{align*}
X_t=m_t^{\star}+\epsilon_t,\quad t\in\mathbb{N},
\end{align*}
where $t\mapsto m_t^{\star}$ evolves linearly and in a sinusoidal fashion while $\{\epsilon_t\}_{t\in\mathbb{N}}$ is an $\mathtt{MA(1)}$ process defined as
\begin{align}\label{ma1}
\epsilon_t=\frac{ \eta_t + a \eta_{t-1} }{\sqrt{1 + a^2}}, \quad t\in\mathbb{N}.  
\end{align}
Here, $a \in \mathbb{R}$ and the elements of the sequence $\{ \eta_t \}_{t \in \mathbb{N}}$ are independent Gaussian random variables with zero mean and unit variance. We then run SES given by Equation \eqref{SES th}, starting from $\hat{m}_0= 8$ and we showcase the tracking trajectory in Figure \ref{fig1}.

\begin{figure}[H]
	\centering
	\begin{subfigure}[c]{0.49\textwidth}
		\includegraphics[width=\linewidth]{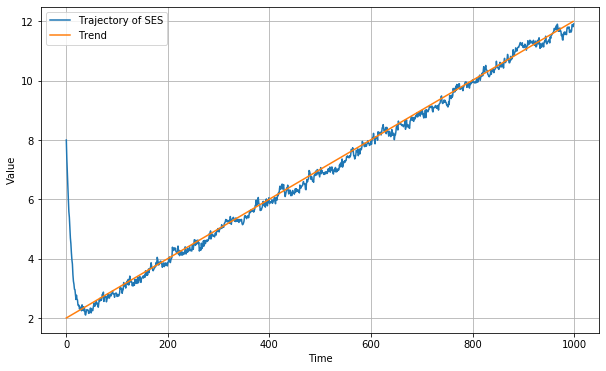}
		\caption{}
		\label{fig2:1a}
	\end{subfigure}
	\begin{subfigure}[c]{0.49\textwidth}
		\includegraphics[width=\linewidth]{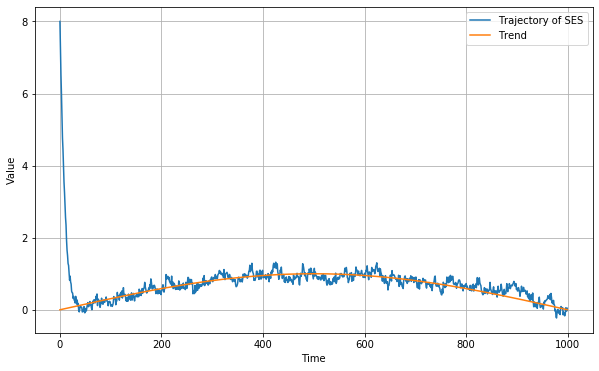}
		\caption{}
		\label{fig2:1b}
	\end{subfigure}
\caption{The evolution of $\{m_t^{\star}\}_{t\in\mathbb{N}}$ in subfigure \ref{fig2:1a} is given by the equation $m_t^{\star} = m_{t-1}^* + 0.1$, starting from $m_0^* = 2$, while $a = -0.4$. The evolution of $\{m_t^{\star}\}_{t\in\mathbb{N}}$ in subfigure \ref{fig2:1b} is given by the equation $m_t^{\star} = \sin( \pi t / 1000)$, starting from $t = 0$, while $a = -0.4$. The SES in both subfigures is iterated $1000$ times with $\hat{m}_0= 8$ and $\alpha = 0.1$.}
\label{fig2}
\end{figure}

The auto-covariance function of the stochastic process defined in \eqref{ma1} is
\begin{align*}
	\gamma(t) = \begin{cases}
			1, & \text{if $t = 0$};\\
            \frac{a}{1+a^2}, & \text{if $|t| = 1$}; \\
            0, & \text{if $|t| \ge 1$};
		 \end{cases}
\end{align*}
thus choosing $a < 0$ we obtain a negative covariance that should aid convergence as implied by Theorem \ref{t1}. In Figure \ref{fig2} we run the SES with $a = -0.4$.
Comparing Figures \ref{fig1} and \ref{fig2} we notice that the neighborhood to which SES converges is smaller in the case of negative covariance, as expected.

\subsection{The autoregressive case}

\begin{figure}[H]
	\centering
	\begin{subfigure}[c]{0.49\textwidth}
		\includegraphics[width=\linewidth]{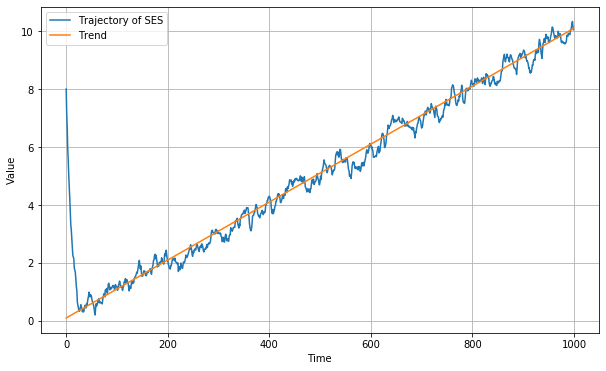}
		\caption{}
		\label{fig:3a}
	\end{subfigure}
	\begin{subfigure}[c]{0.49\textwidth}
		\includegraphics[width=\linewidth]{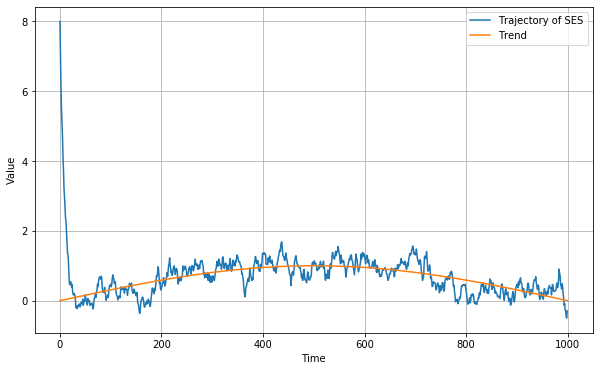}
		\caption{}
		\label{fig:3b}
	\end{subfigure}
\caption{ The evolution of $\{m_t^{\star}\}_{t\in\mathbb{N}}$ in subfigure \ref{fig:3a} is given by the equation $m_t^{\star} = 0.1 + 0.01 t$, starting from $t=0$, while $\theta = 0.2$. The evolution of $\{m_t^{\star}\}_{t\in\mathbb{N}}$ in subfigure \ref{fig:3b} is given by the equation $m_t^{\star} = \sin( \pi t / 1000)$, starting from $t = 0$, while $\theta = 0.2$. The SES is iterated $1000$ times with $\hat{m}_0= 8$ and $\alpha = 0.1$.}
\label{fig3}
\end{figure}

We simulate observations of the form
\begin{align*}
X_t=m_t^{\star}+\epsilon_t,\quad t\in\mathbb{N},
\end{align*}
where $t\mapsto m_t^{\star}$ evolves linearly and in a sinusoidal fashion while $\{\epsilon_t\}_{t\in\mathbb{N}}$ is an $\mathtt{AR(1)}$ process defined as
\begin{align} \label{ar1}
\epsilon_{t+1} = \theta \epsilon_{t} + \eta_t,\quad t\in\mathbb{N} 
\end{align}
where $\theta \in ]0,1[$ and the sequence $\{\eta_t\}_{t\in\mathbb{N}}$ is made of independent Gaussian random variables with  mean zero and unit variance. The auto-covariance function of the stochastic process defined in \eqref{ar1} is
 $$\gamma(t)=   \theta^{|t|} \gamma(0),\quad t\in\mathbb{N}.$$ \\
In Figure \ref{fig3} we showcase the results.

\newpage

\section{Proof of the main result} \label{sec3}

To ease the notation we set $\beta \coloneqq  1- \alpha$ and notice that equation \eqref{SES th} admits the solution
\begin{align} \label{fun}
	\hat{m}_t = \beta^{t-1} \hat{m}_1 + \alpha \sum_{j =1}^{t-1} \beta^{t-1-j} X_j,\quad t\in\mathbb{N}.
\end{align}	
On the other hand, subtracting $m_t^{\star}$ from both sides of equation \eqref{SES th} we get
\begin{align}  \label{start}
	\hat{m}_{t+1} - m_t^{\star} = \beta( \hat{m}_t - m_t^{\star}) +\alpha (  X_t  -m_t^{\star}),
\end{align}	
and taking squares we obtain
\begin{align} \label{square}
	(\hat{m}_{t+1}-m_t^{\star} )^2 = \beta^2 (\hat{m}_t-m_t^{\star})^2 + 2\alpha \beta ( \hat{m}_t - m_t^{\star}  ) (   X_t  -m_t^{\star} ) + \alpha^2 ( X_t  -m_t^{\star} )^2.
\end{align}
We now focus on the product $( \hat{m}_t - m_t^{\star}  ) (   X_t  -m_t^{\star} )$ and rewrite it with the help of \eqref{fun} as 
\begin{align}\label{doublep}
 ( \hat{m}_t - m_t^{\star} ) (   X_t  -m_t^{\star}  ) =&
  \hat{m}_t  (   X_t  -m_t^{\star}  ) -  m_t^{\star}  (   X_t  -m_t^{\star}  ) \nonumber\\
   = & \left(  \beta^{t-1} \hat{m}_1 + \alpha \sum_{j =1}^{t-1} \beta^{t-1-j} X_j \right) (   X_t  -m_t^{\star}  ) -  m_t^{\star}  (   X_t  -m_t^{\star}  ) \notag \\ 
    = &\beta^{t-1}  \hat{m}_1 ( X_t - m_t^{\star} ) +	  \alpha \sum_{j =1}^{t-1} \beta^{t-1-j} (X_j - m_j^{\star}) (   X_t  -m_t^{\star}  ) \notag \\ 
    & +  \alpha \sum_{j =1}^{t-1} \beta^{t-1-j}  m_j^{\star} (   X_t  -m_t^{\star} ) -  m_t^{\star}  (   X_t  -m_t^{\star}  ).
\end{align}
A substitution of the last member of \eqref{doublep} into equation \eqref{square} gives
\begin{align*}
	(\hat{m}_{t+1}-m_t^{\star} )^2 =& \beta^2 (\hat{m}_t-m_t^{\star})^2 + 2\alpha \beta^{t}  \hat{m}_1 ( X_t - m_t^{\star} ) \\
	&+	2 \alpha^2\beta \sum_{j =1}^{t-1} \beta^{t-1-j} (X_j - m_j^{\star}) (   X_t  -m_t^{\star}  )  \\ 
	& +  2 \alpha^2\beta \sum_{j =1}^{t-1} \beta^{t-1-j}  m_j^{\star} (   X_t  -m_t^{\star} ) \\
	&- 2\alpha\beta m_t^{\star}  (   X_t  -m_t^{\star}  ) \\
	&+ \alpha^2 ( X_t  -m_t^{\star} )^2;
\end{align*}
if we now take the expectation of both sides above, recalling that $\mathbb{E}[X_t]=m_t^{\star}$ and $\mathtt{cov}(X_t,X_s)=\gamma(t-s)$ for all $t,s\in\mathbb{N}$, we get
\begin{align*} 
	\mathbb{E} [(\hat{m}_{t+1}-m_t^{\star})^2 ] = \beta^2 \mathbb{E}[(\hat{m}_t-m_t^{\star})^2 ] + 2\alpha^2 \sum_{j =1}^{t-1} \beta^{t-j} \gamma(t-j)  + \alpha^2 \gamma(0).
\end{align*}
or equivalently,
\begin{align}\label{recursive} 
	\mathbb{E} [(\hat{m}_{t+1}-m_t^{\star})^2 ] = \beta^2 \mathbb{E}[(\hat{m}_t-m_t^{\star})^2 ] + 2\alpha^2 \sum_{k=0 }^{t-1} \beta^{k} \gamma(k)  -\alpha^2 \gamma(0).
\end{align}
Here we added the term corresponding to $j=t$ and renamed $t-j$ as $k$.  

The next step consists in rewriting \eqref{recursive} as a recursive equation for $t\mapsto \mathbb{E} [(\hat{m}_{t+1}-m_t^{\star})^2 ]$. First of all, we observe that
\begin{align*} 
\mathbb{E}[(\hat{m}_t-m_t^{\star})^2  ] &= \mathbb{E}[(\hat{m}_t- m_{t-1}^* + m_{t-1}^* -m_t^{\star})^2  ] \\ &=  \mathbb{E}[(\hat{m}_t-m_{t-1}^*)^2  ] + ( m_t^{\star} -m_{t-1}^*  )^2 - 2 ( m_t^{\star} -m_{t-1}^*  ) \mathbb{E}[\hat{m}_t-m_{t-1}^* ];
\end{align*}
thus, utilizing the notation 
\begin{align*}
 D_{t+1} \coloneqq \mathbb{E} [(\hat{m}_{t+1}-m_t^{\star} )^2 ],\quad v_{t+1} \coloneqq \mathbb{E} [\hat{m}_{t+1}-m_t^{\star} ]\quad\mbox{ and }\quad K_t \coloneqq m_t^{\star} - m_{t-1}^*
 \end{align*}
we can reformulate equation \eqref{recursive} as 
\begin{align} \label{t23}
D_{t+1} = \beta^2 (   D_t + K_t^2 - 2  K_t  v_{t}  )+  2\alpha^2 \sum_{k =0}^{t-1} \beta^{k} \gamma(k)  - \alpha^2 \gamma(0) .
\end{align}
Notice that exploiting \eqref{start} one can easily see that the sequence $\{v_t\}_{t\in\mathbb{N}}$ solves the recursive equation
\begin{align} 
	\begin{cases}\label{t22}
		v_{t+1}= \beta v_{t} - \beta K_t,\quad t\in\mathbb{N};\\
		v_1=0;
		\end{cases}
\end{align}	
where we have set $m_0^{\star}= m_1^{\star}$ to obtain
\begin{align*}
v_1=\mathbb{E}[\hat{m}_1- m_0^{\star}]=\mathbb{E}[X_1- m_1^{\star}]=0.
\end{align*}
In fact, 
\begin{align*} 
		\hat{m}_{t+1}-m_t^{\star} &= \beta (\hat{m}_{t}-m_t^{\star}) + \alpha ( X_t - m^*_t)\\
		& = \beta (\hat{m}_{t}- m_{t-1}^* + m_{t-1}^* - m_t^{\star}) + \alpha ( X_t - m^*_t),
\end{align*}
and upon taking expectations of both sides we obtain equation \eqref{t22}. Therefore, the solution to \eqref{t22} takes the form
\begin{align*} 
v_t = - \sum_{h = 1}^{t-1} \beta^{t-h} K_h,\quad t\in\mathbb{N}
\end{align*}
and substituting this expression in \eqref{t23} gives
 \begin{align*} 
 D_{t+1} &= \beta^2 \left[   D_t +  K_t^2  + 2  K_t   \sum_{h = 1}^{t-1} \beta^{t-h} K_h    \right]+   2\alpha^2 \sum_{k =0}^{t-1} \beta^{k} \gamma(k)  - \alpha^2 \gamma(0) \\ &
= \beta^2    D_t + \beta^2 \left[ \left( K_t +  \sum_{h = 1}^{t-1} \beta^{t-h} K_h \right)^2 - \left( \sum_{h = 1}^{t-1} \beta^{t-h} K_h \right)^2    \right] +  2\alpha^2 \sum_{k =0}^{t-1} \beta^{k} \gamma(k)  - \alpha^2 \gamma(0)  \\ &
=\beta^2    D_t + \beta^{2+2t} \left[ \left( \beta^{-t} K_t +  \sum_{h = 1}^{t-1} \beta^{-h} K_h \right)^2 - \left( \sum_{h = 1}^{t-1} \beta^{-h} K_h  \right)^2    \right] +  2\alpha^2 \sum_{k =0}^{t-1} \beta^{k} \gamma(k)  - \alpha^2 \gamma(0)  \\ &
= \beta^2    D_t + \beta^{2+2t} \left[ c_t^2 - c_{t-1}^2 \right] +  2\alpha^2 \sum_{k =0}^{t-1} \beta^{k} \gamma(k)  - \alpha^2 \gamma(0),
\end{align*}
where we have set 
\begin{align*}
c_t :=  \beta^{-t} K_t +  \sum_{h = 1}^{t-1} \beta^{-h} K_h = \sum_{h = 1}^{t} \beta^{-h} K_h,\quad t\in\mathbb{N}. 
\end{align*}
The preceding computation shows that $\{D_t\}_{t\in\mathbb{N}}$ solves
 \begin{align*} 
\begin{cases} D_{t+1} =
\beta^2    D_t + \beta^{2+2t} \left[ c_t^2 - c_{t-1}^2 \right] +  2\alpha^2 \sum_{k =0}^{t-1} \beta^{k} \gamma(k)  - \alpha^2 \gamma(0),\quad t\in\mathbb{N}; \\ 
D_1 = 0, 
\end{cases}
\end{align*}
whose solution can be represented as
 \begin{align*} 
 D_t =  \beta^{2t}  c_{t-1} ^2+\alpha^2 \sum_{i=0}^{t-1} \beta^{2i} \left(    2 \sum_{k =0}^{t-i-2} \beta^{k} \gamma(k)  -  \gamma(0)  \right),\quad t\in\mathbb{N},
\end{align*}
or equivalently,
 \begin{align}\label{b} 
 D_t =   \beta^{2t}  c_{t-1} ^2+2\alpha^2 \sum_{k=0}^{t-2} \gamma(k)     \sum_{i =0}^{t-2-k} \beta^{2i+k}   -  \alpha^2 \gamma(0) \sum_{i=0}^{t-1} \beta^{2i}.
\end{align}

The last step of the proof consists in obtaining an upper bound for
\begin{align*}
	\limsup_{t\to+\infty}\mathbb{E} [(\hat{m}_{t+1}-m_t^{\star} )^2 ]=\limsup_{t\to+\infty} D_{t+1};
\end{align*}
to this aim we now study the asymptotic behavior of the terms appearing in the right hand side of \eqref{b}. 
Recalling that $c_t  := \sum_{h = 1}^{t} \beta^{-h} K_h $ and $K_t := m_t^{\star} - m_{t-1}^*$, with the help of Assumption \ref{ltvp} we can write
\begin{align*} 
	|c_t|  \le K \sum_{h = 1}^{t} \beta^{-h} = K \left( \frac{1-\beta^{-t-1}}{1- \beta^{-1}} -1 \right),
\end{align*}
and hence
\begin{align*} 
	\limsup_{t\to+\infty}\beta^{2(t+1)} c_{t}^2 \le \limsup_{t\to+\infty}K^2 \left( \frac{\beta^{t+1} - 1}{1- \beta^{-1}} - \beta^{t+1} \right)^2 =K^2\frac{\beta^2}{\alpha^2}.
\end{align*}
A simple algebraic manipulation on the second term in \eqref{b} gives
\begin{align*}
	2\alpha^2 \sum_{k=0}^{t-2} \gamma(k)     \sum_{i =0}^{t-2-k} \beta^{2i+k}=\frac{2\alpha}{2 - \alpha} \left[ \sum_{k=0}^{t-2} \gamma(k) \beta^k - \beta^{2t-2} \sum_{k=0}^{t-2} \gamma(k) \beta^{-k} \right];
\end{align*} 
notice that
  \begin{align*} 
\beta^{2t-2} \sum_{k=0}^{t-2} \gamma(k) \beta^{-k} = \beta^t \sum_{k=0}^{t-2} \gamma(k) \beta^{t-2-k} = \beta^t \sum_{j=0}^{t-2} \gamma(t-2-j) \beta^{j},
\end{align*}
and hence from the Cauchy-Schwarz inequality we can conclude that 
  \begin{align*} 
  	\limsup_{t\to+\infty}\left|\beta^{2t-2} \sum_{k=0}^{t-2} \gamma(k) \beta^{-k}  \right|=&  
    \limsup_{t\to+\infty}\left|\beta^t\sum_{j=0}^{t-2}  \gamma(t-2-j) \beta^{j} \right|  \\
       \leq& \limsup_{t\to+\infty}\beta^t\sum_{j=0}^{t-2}  |\gamma(t-2-j)| \beta^{j}  \\    
      \le & \gamma(0) \limsup_{t\to+\infty}\beta^t\sum_{j=0}^{t-2} \beta^{j}
      =0.
   \end{align*}
Therefore, for the second term in \eqref{b} we can write
\begin{align*}
	\limsup_{t\to+\infty}2\alpha^2 \sum_{k=0}^{t-2} \gamma(k)     \sum_{i =0}^{t-2-k} \beta^{2i+k}=\frac{2\alpha}{2 - \alpha} \sum_{k\geq 0}\gamma(k) \beta^k.
\end{align*}
Thus we may conclude that 
 \begin{align*} 
 	\limsup_{t\to+\infty}\mathbb{E} [(\hat{m}_{t+1}-m_t^{\star} )^2 ] \leq \frac{ \alpha }{ 2-\alpha } \gamma(0) + \frac{2\alpha}{2 - \alpha}  \sum_{k\geq 1} \gamma(k) \beta^k       +\frac{\beta^2}{\alpha^2} K^2.
\end{align*}

\bigskip

\bibliographystyle{apalike}
\bibliography{Smoothing_arxiv}

\end{document}